# S-matrix absolute optimization method for a perfect vertical waveguide grating coupler

**Anat Demeter and Shlomo Ruschin***

*Department of Physical Electronics, School of Electrical Engineering Faculty of Engineering,*

*Tel-Aviv University,Tel-Aviv 69978 Israel*

[*]*Corresponding author: ruschin@eng.tau.ac.il*

*Vertical coupling using a diffraction grating is a convenient way to couple light into an optical waveguide. Several optimization approaches have been suggested in order to design such a coupler; however, most of them are implemented using algorithm-based modelling. In this paper, we suggest an intuitive method based on S-matrix formalism for analytically optimize 3-port vertical grating coupler devices. The suggested method is general and can be applied to any 3-port coupler device in order to achieve an optimal design based on user constrains. The simplicity of the model allows reduction of the optimization to two variables and the location of an absolute optimal operation point in a 2D contour map. Accordingly, in an ideal device near 100% coupling efficiency and insignificant return loss could be achieved. Our model results show good agreement with numerical FDTD simulations and can predict the general tendencies and sensitivities of the device's behavior to changes in design parameters. We further apply our model to a previously reported high contrast uni-directional grating coupler device and show that additional improvement in the coupling efficiency is achievable for that layout.*

# 1. Introduction

Vertical grating couplers are an attractive option for coupling light into an optical waveguide due to their high performance, their easy fabrication, and their reduction of the system's tuning complexity [1–4]. In these gratings, the ability to couple light into a specific direction is necessary since the coupling is symmetrical, transmitting light both to the left and to the right directions. A critical issue, which can significantly affect the source functionality of this configuration, is the back-reflection from the grating into the incoming source direction.

In recent years, many studies have been done in order to optimize the design and fabrication of these vertical couplers [5–8]. However, most of the methods suggested so far are implemented by complex algorithms and numerical simulations that are difficult for the user to follow and reproduce. This problem motivated us to develop a model, which is intuitive, based on basic physical assumptions and easy to implement and realize. It helps the user to understand whether further improvement in a specific design can be achieved without the use of any complex machine learning or data processing algorithms.

In this paper, we perform a comprehensive analysis of an absolute optimization method for vertical grating coupling based on S-Matrix formalism and its corresponding constraints of reciprocity and power conservation. We show that, in spite of these ideality assumptions, our results are in good agreement with numerical FDTD simulations, and able to predict the general tendencies and optimal parameters of the device. Using our model, an absolute upper limit for the efficiency and lower limit for the return loss are determined. By locating a simulated practical device on a contour map and adjusting its position, the device efficiency can be further improved. Furthermore, mechanical tolerances and optical bandwidth can be straightforwardly evaluated. As an illustration, we analyze a device reported recently [5] and suggest a slightly modified version with a fairly enhanced performance. For a more general validation of our model accuracy, we compared between the analytical and numerical simulations pattern changes for several grating coupler designs, and received very good correlation between these approaches.

## 2. S-matrix model

The proposed optimization method is general and relevant to any 3-port coupler device. It is based on scattering matrix model which was previously applied to design a grating-coupled laser resonator [9]. In our case it assumes a 3-port vertical grating coupler, meaning it was designed to allow only negligible light transmission into the substrate [2,6,10,11]. A schematic drawing of the device is presented in Fig.1.

The S-matrix is defined by the following expression [12]:

$$S = \begin{pmatrix} r_{11} & t_{12} & t_{13} \\ t_{21} & r_{22} & t_{23} \\ t_{31} & t_{32} & r_{33} \end{pmatrix}, \tag{1}$$

where $r_{ij}$ and $t_{ij}$ ($i,j$=1,2,3) are the reflection and transmission coefficients respectively. In our model the vertical grating coupler is symmetric, meaning ports 1 and port 3 (Fig.1) are identical i.e. $t_{12} \equiv t_{32}$. As a result and by further requesting reciprocity, the reflection and transmission magnitudes coefficients can be represented by four independent variables:

$$\sigma \equiv |t_{12}| = |t_{21}| = |t_{23}| = |t_{32}|, \ \eta \equiv |t_{13}| = |t_{31}|, \ \xi \equiv |r_{11}| = |r_{33}|, \ \rho \equiv |r_{22}|. \tag{2}$$

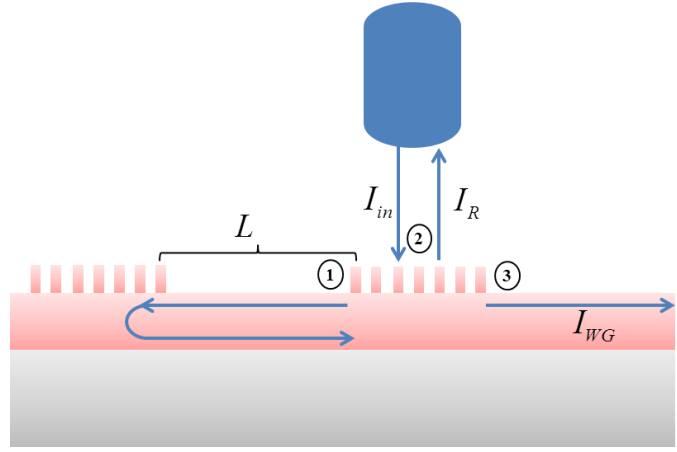

Fig.1. Schematic drawing of the suggested device. The numbers in the circles represent the ports numbers, the waveguide length is $L$, and $I_{in}, I_R$ and $I_{WG}$ are the input intensity and output calculated intensities respectively.

Using this matrix and assuming a steady state regime, the back-reflected field into the optical fiber and the right-propagating field in the waveguide are calculated:

$$E_R = a_1 t_{12} + a_2 r_{22} = a_1 \sigma \exp(i\phi_\sigma) + a_2 \rho \exp(i\phi_\rho),$$
$$E_{WG} = a_1 t_{13} + a_2 t_{23} = a_1 \eta \exp(i\phi_\eta) + a_2 \sigma \exp(i\phi_\sigma). \quad (3)$$

Both fields are the sum of complex field amplitudes multiplied by the reflection and transmission coefficients of the grating defined above. $a_1$ is the complex field amplitude of the propagating light towards the coupling grating from the right and $a_2$ is the input field amplitude ($|a_2|^2 = I_{in}$). $\phi_\sigma$ and $\phi_\xi$ depend on the four mentioned magnitudes as defined in [9]. Assuming steady state condition, we derive the following relation between these two complex field amplitudes:

$$a_1 = \frac{\sigma \exp(i\phi_\sigma) r_G \exp(-2i\beta L)}{1 - r_G(\xi \exp(i\phi_\xi))\exp(-2i\beta L)} a_2, \quad (4)$$

where $L$ is the waveguide length, $\beta$ is the light propagation constant inside the waveguide, and $r_G$ is the waveguide left reflector reflectivity.

From the unitary condition of *S* the following power conservation relations are accepted:

$$\rho^2 + 2\sigma^2 = 1,$$
$$\eta^2 + \sigma^2 + \xi^2 = 1. \tag{5}$$

Using Eq.(5) the problem can be further reduced into two independent variables, namely $\rho$ and $\eta$. $I_R$ and $I_{WG}$, can then be easily calculated. By plotting a contour map of $I_R$, the optimal grating coefficients $\rho$ and $\eta$ can be extracted. For each grating design, there is thus a unique optimal set of values for the grating diffraction efficiencies $\rho^2, \eta^2, \sigma^2, \xi^2$ and length L, as explained below.

As $\rho$ and $\eta$ run between their entire value range, {0-1}, the map furnishes the absolute optimal back-reflection and waveguide length. Although these optimal values will in general be unattainable, in a practical device they can provide a benchmark for a given design, and more important, suggest a path direction for attempting its improvement. In the following section we demonstrate that procedure for a specific design found in the recent literature [5].

## 3. Optimization Results

Our optimization approach compares between both analytical and numerical simulations. The numerical simulations were performed using Lumerical FDTD solutions software. We demonstrate our model implementation by means of two examples:

1. A general and ideal directional vertical grating coupler design, including a side back-reflector, for calculating the device theoretical efficiency limits.

2. A specific high contrast grating (HCG) vertical coupler device for indicating how our model can improve a previously reported design [5].

In addition, we compare between the analytical model and the numerical simulations results for several vertical grating couplers designs with different sets of $\rho$ and $\eta$ values.

Referring back to Fig.1, the grating structure excites the fundamental TE mode in the waveguide at a $\lambda=1.55\mu m$ wavelength. The mode's effective refractive index is 2.12, corresponding to a 100nm thick silicon-on-isolator waveguide. An ideal waveguide reflector was assumed at the left side of the device with reflection coefficient $r_G=1$. We started the optimization by producing contour plots of the device using our S-matrix analytical model algorithm. From the requirement of real grating phase arguments one deduces boundary limits for $\rho$ and $\eta$ that restricts the contour maps to a specific allowed region [9]:

$$\frac{(1-\rho)}{2} \leq \eta \leq \frac{(1+\rho)}{2}. \tag{6}$$

These limits are fundamental and define the working area within a grating coupler can be designed. In Fig. 2(a), a contour plot for the back reflected intensity as function of the independent variables $\rho$ and $\eta$ is displayed. For each value of these parameters, the optimal value of $L$ was established and plotted as contours in Fig. 2(b). $\rho$ and $\eta$ were allowed all possible values in their definition range, and therefore, the optimal values extracted from the plots of Fig.2 are absolute and become fundamental limits of performance of all 3-port couplers, regardless of the specific geometrical details of the coupler. The optimal waveguide length, $L_{opt}$, was assumed to be in the range of $0<L<1\mu m$, However due to the periodicity in the resonator's length, the same back-reflection result will be obtained for any length $L'_{opt}$ satisfying the condition:

$$L'_{opt} = L_{opt} + \frac{\lambda}{2 \cdot n_{eff}} \cdot m \ , \ m=1,2,3,... \tag{7}$$

where $\lambda$ is the input source wavelength and $n_{eff}$ is the waveguide effective refractive index. An absolute optimal effective distance of $L_{opt}=0.8077\mu m$ and a minimal normalized back-reflection intensity of $3.13 \cdot 10^{-8}$ (-75dB) were determined for grating coupler efficiency amplitudes $\rho = 0.4809, \eta = 0.62$. The minimal back-reflection value is practically zero and the small back-reflection intensity was limited by the numerical sampling subdivision of $\rho$ and $\eta$. In practical terms: in order to determine the optimal design, first the optimal back-reflected intensity and the matching $\rho$ and $\eta$ parameters were extracted from Fig.2.(a), and then the waveguide optimal length was determined from the contour map in Fig.2.(b). The remaining transmission coefficients, σ and ξ, are obtained from Eq. (5). Since the model assumes power conservation, the uni-directional waveguide coupling intensity is calculated by the relation: $I_{WG}(\rho,\eta) = I_{in} - I_{R}(\rho,\eta)$. The model allows also straightforward calculation of tolerances and the optimal device shows a large bandwidth acceptance range: for the entire telecom C-band, i.e wavelength from 1530 to1565nm, the back-reflection is less than -20dB as illustrated in Fig.2.(c).

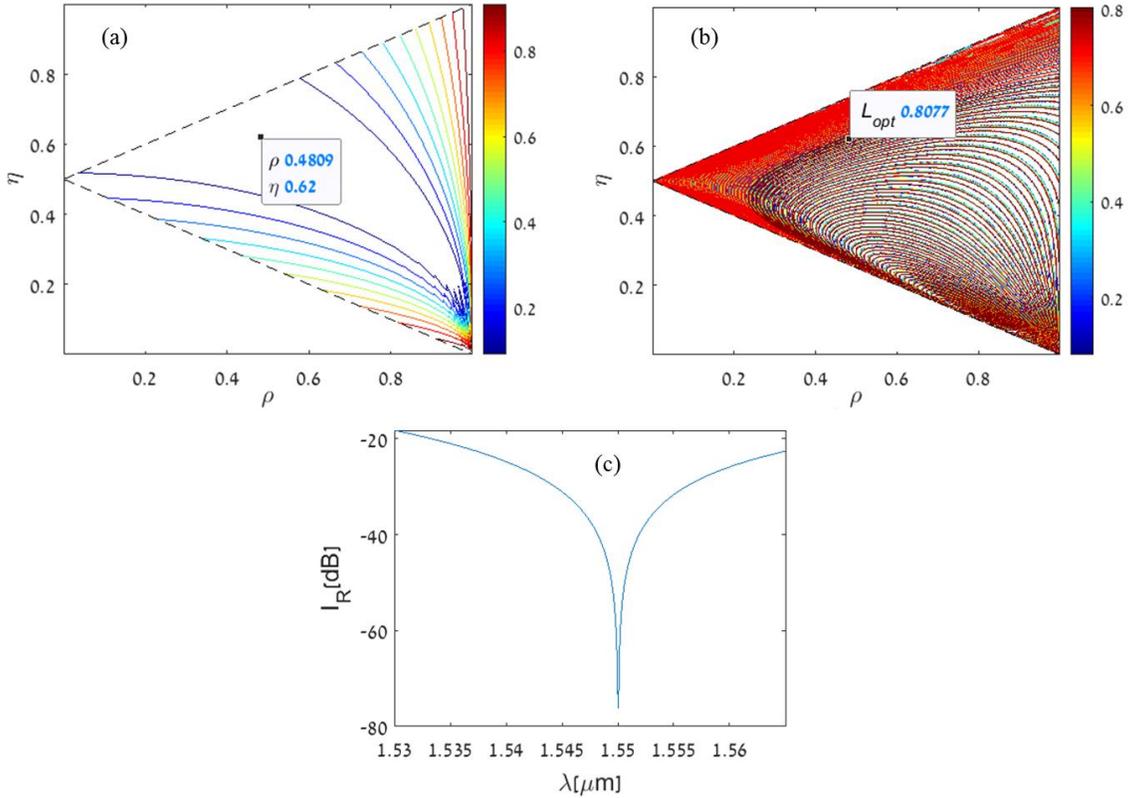

Fig.2. Contours plots of (a) the back-reflected intensity and (b) the waveguide length, as a function of the two independent efficiencies parameters $\rho, \eta$. The optimal values are marked on both figures. In (c) the back-reflected intensity as function of wavelength in the telecommunication C-band range is plotted.

Results insofar refer as stated to an ideal lossless system. For demonstrating how our S-matrix model can be also useful in a practical situation, we chose a previously reported device, namely the HCG coupler with low transmission to the substrate which was suggested by Zhu et al. [5]. In Fig.3 the device electrical power density and the contour map for the waveguide right-directed coupling intensity are plotted. We simulated numerically their design and extracted the relevant $\rho$ and $\eta$ parameters. Then we located the matching coordinate of the device ($\rho = 0.3839, \eta = 0.5742$), on the intensity contour plot in Fig.3.(a). and reproduced the reported 88% coupling efficiency. However, this efficiency value is not the optimal one since it is not located on the highest contour line. In order to get improved result we slightly changed the grating coupler design parameters namely: duty cycle=0.59, grating height=950nm, period=0.733nm and 250nm gap which lead to different $\rho$ and $\eta$ values. A directed coupling efficiency of 91% was attained for the values $\rho = 0.4631, \eta = 0.5995$. As seen, the improved device coordinate ($\rho, \eta$) is matched to higher contour line in Fig.3.(a), meaning an improvement of 3% in the coupling efficiency. This small improvement is not negligible since it transduces into lower return losses, where tolerances are more stringent.

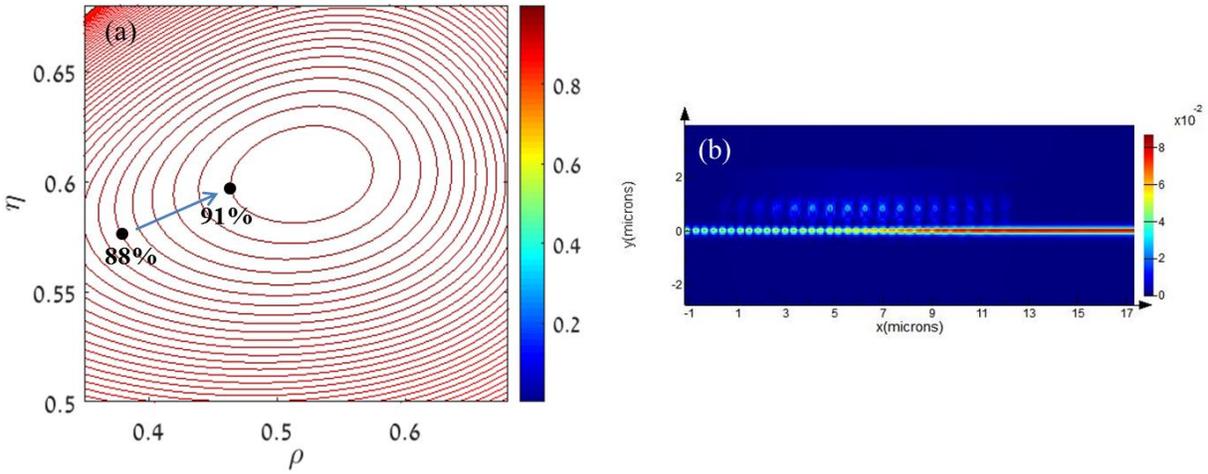

Fig.3. Plots for the simulated HCG directional vertical coupler device: (a) counter plot of the directional coupling efficiency of the device using our S-matrix model. 88% coupling efficiency was reported in [5] and 91% coupling efficiency was simulated by us for the improved design. In (b) the electrical power density of the device is displayed.

The S-matrix model assumes ideal conditions, such as power conservation and reciprocity. However in reality this is not the case. There is energy loss when coupling from the fiber to the waveguide and vice versa caused by mode mismatch. In addition the reciprocity is also not well kept because of different beam profiles in the fiber and in the optical device. Another issue that can affect our model results is the fact that it is based on 3-port grating coupler design, where in practice there will be finite transmission into the substrate [2,5]. In spite of these approximations, our analytical model followed the numerical simulations accurately and guided us to achieve an improved design of the coupler. To further validate this assumption we numerically simulated several directional HCG vertical couplers designs, starting with the coupler reported in [5]. Each grating design had slightly different parameters (duty cycle and height), and for each design we numerically determined $\rho$ and $\eta$ values and the right-directed waveguide coupling efficiency. We then inserted the numerically measured $\rho$ and $\eta$ set of values to the analytical S-matrix model and calculated the ideal efficiency for each $\rho,\eta$ pair. We plotted two graphs: one for the right-directed coupling intensity as a function of $\rho$ and the other as function of $\eta$ (Fig.4). As it can be seen from the two plots, the intensity changes in the analytical and the numerical simulations are very well correlated and the analytical model properly follows the numerical simulations pattern. The gap between the intensity values in the analytical and numerical simulations is caused by the power loss mechanisms mentioned above. As seen, even local maxima and minima are very well correlated for both models.

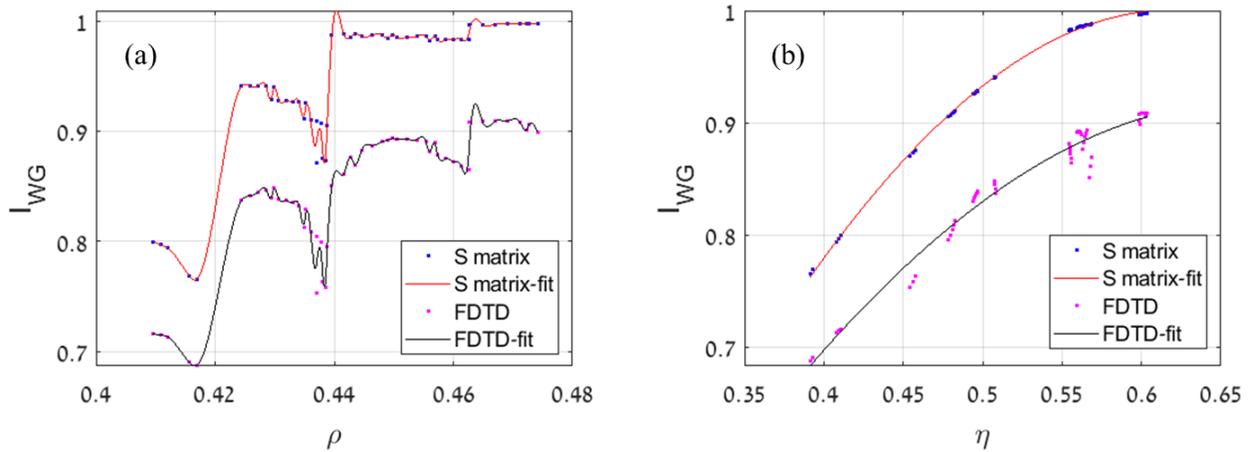

Fig.4. Plots for the right directed coupling intensity as function of (a)$\rho$ and (b)$\eta$. The dots in each plot represents the calculated intensity values using the S-matrix analytical model (magenta) and the numerical FDTD (blue) simulations. Line MATLAB fits for both cases are also shown.

## 4. Conclusions

We presented a solvable method based on S-matrix formalism for theoretically determining the absolute optimal coupling conditions for vertical coupling using grating arrangement aided by back-reflection. Being determined by basic principles of power conservation and reciprocity, the optimal values must be considered as upper limits of attainable efficiencies. Applied to a specific practical example, the model guided us into a further improvement in the coupling efficiency. The ability of this simple model to predict general tendencies of a 3-port couplers, should facilitate its implementation in further configurations for optical communication or sensing [13].